# IMPROVING FEEDBACK FROM AUTOMATED REVIEWS OF STUDENT SPREADSHEETS

## *Working Paper*

Sören Aguirre Reid, Frank Kammer, Jonas-Ian Kuche, Pia-Doreen Ritzke, Markus Siepermann, Max Stephan, Armin Wagenknecht

Technische Hochschule Mittelhessen, Wiesenstraße 14, 35390 Gießen, Germany

## Abstract

*Spreadsheets are one of the most widely used tools for end users. As a result, spreadsheets such as Excel are now included in many curricula. However, digital solutions for assessing spreadsheet assignments are still scarce in the teaching context. Therefore, we have developed an Intelligent Tutoring System (ITS) to review students' Excel submissions and provide individualized feedback automatically. Although the lecturer only needs to provide one reference solution, the students' submissions are analyzed automatically in several ways: value matching, detailed analysis of the formulas, and quality assessment of the solution. To take the students' learning level into account, we have developed feedback levels for an ITS that provide gradually more information about the error by using one of the different analyses. Feedback at a higher level has been shown to lead to a higher percentage of correct submissions and was also perceived as well understandable and helpful by the students.*

## 1 Introduction

Spreadsheets are the most popular end-user programming environment and therefore widely deployed in organizations for computational or decision-making processes (Abraham and Erwig, 2007a; Barowy et al., 2018; Jansen and Hermans, 2021; Mukhtar et al., 2022). There are many reasons for the popularity of spreadsheets. First, spreadsheets allow easy calculation, evaluation, and data analysis. Second, data management in spreadsheets is simpler, faster, and easier than in a database. Third, the intuitive user interface of spreadsheets makes it easy for users to copy cells or to drag on a cell to fill a column. So, users do not have to be IT experts to use spreadsheets. Not surprisingly, many reports of spreadsheet errors (e.g., financial losses for organizations; spreadsheet error led to Edinburgh hospital opening delay) appear in the global media at a fairly consistent rate (Eusprig, 2023). For this reason, it is critical to support spreadsheet users with feedback at the beginning of their learning process so that they can develop into professional users.

Several studies followed this motivation and developed approaches to assist spreadsheet users in spreadsheet debugging in the business context (e.g., Abraham and Erwig, 2007a; Chamber and Erwig, 2010; Joharizadeh, 2015; Jansen and Hermans, 2021; Jannach et al., 2014; Mukhtar et al., 2022). These tools help users to identify cells in cluttered tables that are likely to be defective. For instance, studies that focused on spreadsheet debugging utilized reference counting to detect errors in the spreadsheet (Joharizadeh, 2015) or used a block-based view for Excel formulas to make them easier to understand (Jansen and Hermans, 2021). Further studies focused on detecting spreadsheet smells. Spreadsheet smells indicate, for example, missing input values, wrong numeric values, or mistyped strings (Koch et al., 2019a). For detecting spreadsheets smells, former research performed static analyses to detect spreadsheet smells (Koch et al., 2016) or calculated metrics that detect the smells with respect to size, level of coupling, and the use of functions (Hermans et al., 2012; 2015; Jansen and Hermans, 2015). A few studies focused on the quality of spreadsheets (Cunha et al., 2012; Koch et al., 2021). For instance, they defined a set of specific metrics to measure the spreadsheet characteristics (e.g., functionality, reliability).

Due to the widespread use of spreadsheets in business contexts (e.g., decision-making), many different study programs of universities use spreadsheets intensively in their teaching context (Rafart Serra et al., 2017; 2018; 2021). For example, spreadsheets are used in computer science (database), business science (e.g., controlling, operations research) or media systems courses (business administration and information systems). However, the usage of spreadsheets in the teaching context is still mostly guided by the lecturer and not supported by automated evaluation with feedback based on the submission. Though, using an intelligent tutoring system (ITS) in the teaching context to assess student spreadsheet assignments benefits the student and the lecturer. From the student's perspective, to become proficient in spreadsheet usage practicing is mandatory. But not only practicing matters. When it comes to the correctness of spreadsheets in terms of content (e.g., formulas), students need feedback on what they did right and wrong or how they can improve their submission regarding the quality (Hattie and Timperley, 2007). From the lecturer's perspective, manually examining students' spreadsheets (e.g., spotting formula mistakes) and providing feedback requires careful examination and is very time-consuming (Barowy et al., 2018). Despite the obvious need and arguments for an intelligent tutoring system (ITS) for spreadsheets, only two studies have investigated this in a teaching context.

The studies by Rafart Serra et al. (2017; 2018) developed and introduced an automatic spreadsheet corrector at the university. The automatic spreadsheet corrector uses a reference solution to examine spreadsheet errors. For the analysis of the submitted spreadsheet, they utilize a value-matching approach. In the context of Excel, value matching is the comparison of two values in the cell without taking the underlying formulas into account. Based on the value-matching approach, the automatic spreadsheet corrector generates feedback information for the students (e.g., whether the spreadsheet is right or wrong or the location of the error cell). However, the automatic spreadsheet corrector can only generate feedback if the lecturer has stored it in the cell in the spreadsheet. That means, that for each task, feedback must be created in each of the spreadsheet cells manually.

Thus, although skills in building and using spreadsheets are of highest interest for business, ITS that support spreadsheet exercises and practicing with automatic marking and feedback is still extremely scarce. This paper addresses this gap and contributes to the stream of ITS for spreadsheets in the following ways.

First, we address the understudied area of assisting spreadsheet users in error detection in the teaching context. For this purpose, we developed an Excel checker, and tested it in the teaching context. The goal of the ITS is to prepare students in the best possible way for practice and to enable them to identify spreadsheet errors. In other words, students should develop into self-regulating learners to recognize and correct errors independently (e.g., by thinking or trying them out) (Nicol and Macfarlane-Dick, 2006). To achieve this goal, the ITS automatically marks student submissions based on a reference solution and gives feedback.

Secondly, the aim of the ITS is not to simply provide the solution, students should be guided to the solution. In contrast to existing systems, such as Rafart Serra et al. (2017; 2018; 2021) (see subchapter related work), the ITS supports solution finding with automatic generated feedback. In this case, the lecturer does not have to create feedback for the respective cells in the spreadsheet, instead the ITS generates the feedback. This makes it easier for lecturers to create new assignments and motivates them not to use the same assignment every term. So, the goal of the system is to give the best feedback with the least amount of effort from the lecturer. Furthermore, the ITS supports the student learning process with seven different feedback levels. Each level of feedback reveals more detailed information to support the solution-finding process. The seven feedback levels were derived from the feedback literature and built on each other for an ITS (see chapter feedback levels).

Thirdly, our ITS also uses a reference solution like the automatic spreadsheet corrector (Rafart Serra et al., 2017; 2018), but differs in the analysis method. For the comparison, we use a dependency graph to identify the error cells in the spreadsheets (Hofer et al., 2021; Hofer and Wotawa, 2014). This allows our ITS to identify the location of the error and sort out errors passed on by references (propagated error). In addition, we can analyze the formulas of the error cells and assess the quality of the formulas (e.g., it is preferable to use an AVG formula instead of calculating the average manually) using the dependency graph.

The remainder of this paper is organized as follows: The next section reviews the existing literature in the field of ITS for spreadsheets and highlights the contribution of this paper in more detail. In the following section, we describe the derivation of the seven feedback levels based on the feedback literature and provide an example of the feedback levels. Then, we discuss the implementation of the feedback levels in the ITS for Excel tasks and give a brief overview on how to adapt them for other task types, such as SQL and Java. To evaluate our ITS, we surveyed 62 students regarding their perception of the feedback provided by the ITS. The paper closes with a discussion of the results, the according implications and the limitations of this study.

## 2      Related Work

For the literature review, we developed the relevant key terms for the search string and identified the database (AIS, IEEEXplore, ACM) for our literature review in the first step. In the second step, we run the literature review with the following search strings for published journal papers and conference proceedings between 2000 and 2022: "spreadsheet" AND ("error detection" OR "automatic error detection" OR "code smells") AND ("quality assurance" OR "feedback"). In total, we identified 36 publications. In step three, we consider the following criteria for the title, abstract and full analysis: 1. the system should analyze spreadsheets with a reference solution (e.g., error detection, quality evaluation); 2. students should receive (semi- or automated) feedback (e.g., correct or false). Of these 36 papers, only two papers are relevant for our study. Most of the existing literature is business-oriented and mainly focused on recognizing errors, rather than helping students to resolve them. Also, in a non-education use case, you have no knowledge of the correct solution, but in education, you have a reference solution created by the educator which can be used to improve learning support. In recent years, the educational landscape has shifted towards remote learning. This shift allows for an extension of the capabilities of Intelligent Tutoring Systems. Rafart Serra et al. (2017; 2018; 2021) use a reference solution to examine spreadsheet errors in the teaching context. The first and second studies (Rafart Serra et al., 2017; 2018) developed an automatic spreadsheet corrector for the ACME platform. The automatic spreadsheet corrector compares cells with the corresponding ones in the student spreadsheet. Based on the comparison, the system generates feedback if these values do not match. For the value-matching, the lecturer enters a reference solution (e.g., formulas, predefined range of values) to obtain the results based on the parameter values and the provided information by the submission. Rafart Serra et al. (2017; 2018) not only conducted a value-matching approach but also considered a margin for errors in the implementation (e.g., predefined range of values for rounding issues). Despite the provision of individualized feedback, the lecturer must previously store it in the automatic spreadsheet corrector for each cell in the spreadsheet. This means that the automatic spreadsheet corrector can only provide feedback if the lecturer has once created it for each task. Thus, the automatic spreadsheet corrector only automatically detects errors, but generates no automatic feedback. The first study also investigated the student's perception of the system. The results demonstrated a very good opinion of the learning system. The second study introduced and tested the system at the university. However, no results regarding the student's performance were reported. Considering the feedback, the automatic spreadsheet corrector informs the students whether their spreadsheet is correct or not correct, about the error location and the progress, either textual or numerical (e.g., a mark) or both (Rafart Serra et al. 2017; 2018). Additionally, the automatic spreadsheet corrector also provides information on the task context (e.g., the calculation includes production capacities). The third study (Rafart Serra et al., 2021) only investigated the perception of the feedback and how the system affected the students' performance. In contrast to the studies of Rafart Serra et al. (2017; 2018), the developed ITS provides more detailed feedback (e.g., formula details, formula quality) to the students.

Having a look at the other studies found, which are not situated in the teaching context, most studies investigated end-user spreadsheet debugging in the business context. These studies aim at identifying cells in spreadsheets that are likely to be defective (e.g., Ahmad et al., 2003; Antoniu et al., 2004; Abraham and Erwig, 2007a; Abraham and Erwig, 2007b; Barowy et al., 2014; Jannach and Engler, 2010; Chamber and Erwig, 2010; Hofer and Wotawa, 2014; Joharizadeh, 2015; Barowy et al., 2018; Jansen and Hermans, 2021; Mukhtar et al., 2022). Although not directly related to the teaching context, as they do not revert to a reference solution, some of these papers provide useful methods for the evaluation of spreadsheets in the teaching context.

Hofer and Wotawa (2014) as well as Barowy et al. (2014) both applied a dependency graph approach. Hofer and Wotawa (2014) introduced a dependency-based approach for model-based fault localization in spreadsheets. To achieve this, they combined a value-based and dependency-based technique. The combination overcomes the disadvantages of the value-based approach and dependency-based approach. The hybrid model identifies potential error cells in the spreadsheets. Thus, the user receives a set of error cells, but needs to examine all these cells to find the error cell. Barowy et al. (2014) developed CheckCell to support user debugging in spreadsheets by identifying cells that have a large impact on the spreadsheet. For this reason, they utilized a dependency- graph in the first step to determine the effect of a particular input on the formulas in the computation forest because a formula is a node in a computation tree, and this formula's leaves are input values. The system CheckCell assumes that the value of a function changes significantly when an erroneous input value is corrected. Therefore, the authors considered an impact analysis and impact scoring. Considering the provided feedback by CheckCell, each potential error is presented to the user one-at-a-time and the user must either mark the cell as correct ("Mark as OK") or fix the error ("Fix Error"). The systems of both papers inform the user whether a spreadsheet is correct or not and about the error location. Hence, the dependency graph is a suitable method to identify student mistakes. With the help of a dependency graph, the values of a reference solution can be compared with the values of a student submission. In addition, based on the analysis of dependent cells propagated errors can be found.

Ahmad et al. (2003) and Antoniu et al. (2004) utilized unit testing for detecting user errors in spreadsheets. Ahmad et al. (2003) defined the concept of headers and relationships. They considered a header as a common unit for a group of cells and assumed that all headers are known for the unit checker. Relationships can be distinguished into is-a (instances and subcategories) and has-a (properties of items or sets) relationships. With these, they presented a list of rules that help identify weaknesses in spreadsheets that are likely to be errors. Furthermore, they implemented a unit checker and used the notion of units as basic elements of checking. The feedback provided by the unit checker is limited to tell the user if the spreadsheet is correct or not, the error location, and if there is any error propagation. The second study by Antoniu et al. (2004) developed XeLda, a unit checker tool for validating the unit correctness of spreadsheets. For this purpose, a unit system is defined for value fields based on user-defined labels. For calculation fields, the unit is then calculated based on the references. The program finds either fields with mismatching labels or a formula that uses units inconsistently. Regarding the feedback, XeLda was able to inform the user whether the spreadsheet was correct or not. Moreover, the system flags errors by coloring the cells where the error occurs in the spreadsheet (orange for a match error; yellow for a consistency error). Additionally, XeLda provided information about error propagation and colored purple all cells that depended on the error cell.

Hull and du Boulay presented a system that provides several layers of corrective and metacognitive feedback to guide learners in crafting accurate and efficient SQL queries. They presented six distinct levels of help, starting with simply indicating if the query was correct or not, and progressing to series of steps to guide the user to a complete solution, allowing learners to receive the level of assistance tailored to their individual needs. However, one disadvantage of this system is that it is highly rule-based. This reliance on predefined rules and patterns to detect errors and provide feedback can be effective in providing immediate feedback and guidance, but also can be limiting when handling unique student submissions.

In sum, unit testing can help to extract the information from the header cells of an erroneous cell. With the information from the header cell, related information can be found course material with the additional help of Natural Language Processing approaches.

# 3      Feedback Levels

## 3.1    Development of Feedback Levels

A task in the ITS is set up by a lecturer who uploads an exercise description of an economical or mathematical problem, a reference solution in form of an Excel spreadsheet, and possibly learning materials in PDF format. The students then have to fill their own spreadsheet in and modify one or more worksheets corresponding to the problem. The spreadsheets are usually provided with row and column descriptions as well as input values, so that students must follow a given format. However, the submission does not have to match the structure of the reference solution exactly, only a few important cells need to be in the same position. Furthermore, the formulas used do not have to be the same as the reference solution, they just must produce the same result. Afterwards they can upload their Excel spreadsheet for that task and get immediate feedback on their submission. The feedback can be differentiated into seven different feedback levels (FL) for an ITS. Each feedback level builds on the previous one and helps students to find the solution step by step. The goal of the feedback of the ITS is that students become self-regulated learners. Self-regulated learners are less dependent on external lecturer support (e.g., feedback) and have an incremental view of their abilities, which helps them to overcome difficulties in task solving (Dweck, 1999; Grant and Dweck, 2003). Therefore, the provided amount of feedback in the ITS should help the students to develop an incremental view of their abilities (Nicol and Macfarlane-Dick, 2006). Concerning the feedback content, we follow the "good" feedback principles by Nicol and Macfarlane-Dick (2006) and avoid high-stakes assessments (e.g., marks or grades) as feedback information, which negatively impacts motivation for learning (Harlen and Crick, 2003). Instead, we encourage students' lifelong learning by feedback comments to help the student to master the task and the subject (Dweck, 1999).Regarding the feedback level, already pure corrective feedback, also called knowledge of results (FL 1) (Narciss, 2008), which classifies students' solutions simply as correct or incorrect, can be very useful for students (e.g., Sturges, 1978; Lysakowski and Walberg, 1982; Tenenbaum and Goldring, 1989). But corrective feedback does not give hints concerning the mistakes. Thus, when students get stuck, pure corrective feedback does not direct students to the right way of solution. That is where hints mechanisms come into focus where feedback is given that guides in a certain way through the solution process (Hattie and Timperly, 2007). Hints are the most effective form of feedback when they help students in rejecting erroneous hypotheses and provide direction for searching and strategizing (e.g., Earley, 1988; Harackiewicz, 1979; Harackiewicz et al., 1984; Wood and Bandura, 1987). Therefore, the ITS utilizes hints as feedback to help students to solve the assignments gradually by thinking and trying instead of getting an immediate solution (Early, 1998). Interestingly, this kind of feedback is not further differentiated into subcategories. A reason may be that the plethora of different kinds of exercises are considered as too heterogeneous. This may be true when different concepts of exercises like essays, modeling or calculations are regarded. But as long as one exercise category is investigated, different categories of hints can be differentiated. This paper focuses on calculations in spreadsheets. Hence, there are basic values and mathematical or logical relationships.

Instead of pure corrective feedback, FL 2 provides hints regarding the location of cells, known as knowledge about mistakes (Narciss, 2008; Keuning et al., 2016), where the (evaluated) values are incorrect (e.g., check columns b5 and c5). However, a wrong cell implies that all dependent cells are also wrong and returned as wrong.

FL 3 provides information on the location of cells where the (evaluated) values are incorrect, but if an error is propagated to other cells, only the location of the original error is reported to the student (e.g., in Table 1, "check column D3" as the original error location). Based on FL 3, the students know in which cell the original error is, and therefore the formula must be wrong there. But if the student lacks basic knowledge about the formula concept, they will receive no more information than the information provided by FL 3.

That is where FL 4 comes into focus where the feedback provides information about the so-called knowledge about the concept (Keuning et al., 2016; Narciss, 2008; Narciss et al., 2014) or tutoring feedback to the students (Glouli et al., 2006). At this, the lecturer can annotate certain cells or areas with basic information (e.g., basic information about operators or formula syntax), hints (e.g., remember to calculate the average), or links using an interactive worksheet preview. However, FL 4 does not tell the student what is wrong with the formula. If the students need help discovering a concrete problem, they have to rely on the information from FL 5.

FL 5 provides knowledge about the error (Narciss, 2008; Keuning et al., 2016) and can be divided into two subcategories. FL 5a provides information on whether the formula error is an operator (e.g., +, *) or a function (e.g., AVG, IF Statement). If the student does not complete the task correctly with the help of feedback level 5a, s/he will receive FL 5b next, which provides information whether the mistake concerns a constant (e.g., value or string) or a reference (e.g., range or absolute/relative reference) used. Although the student receives information about what type of error is present, FL 5 does not provide information on how to correct the error.

FL 6 contains information about the knowledge on how to proceed by providing hints on mistakes or explanations for the error correction. This kind of feedback significantly affects students' successful completion of tasks (Narciss et al., 2014). But the feedback only contains relevant information for task processing rather than the correct answer to the task (Keuning et al., 2016; Narciss, 2008). For example, FL 6 tells the student to use the + operator, but not which operator should be changed in the formula. However, FL 6 does not tell the students about the quality of the submission. For instance, it is preferable to use a SUM-formula when you add up more than three cells. Hence, once students have completed the tasks correctly, students will receive feedback on the quality of the submitted solution based on quality criteria (e.g., formula length, general number of formulas) in FL 7.

FL 7 aims to promote reflection, and students should think about their submission (Gouli et al., 2006) by providing information that leads to greater possibilities for learning (Hattie and Timperley, 2007). For this, the system evaluates students' submissions based on quality criteria that promote conciseness and readability of the spreadsheet, like the length of the formula chain, the length of a formula, or the same calculations in different fields (which refers to copy and paste). If, based on those criteria, the submission can be improved, feedback containing suggestions for improvement is provided to the student.

### 3.2 Example of Feedback Levels

To support the understanding of the feedback levels, an example is provided below. In the example, two exams (Ex.) are used to determine the grade for three students. First, the students have to calculate the average grade for each exam. The next step is to determine the final grade, which is the average of the two exams. Table 1 shows the spreadsheet of the example, containing several errors compared to the solution in Table 2. Both tables are divided into two parts, the left side shows the values of the spreadsheet, and the right side shows the underlying formulas. Each erroneous value or formular is highlighted in red. In detail, the cells D3 and C6 contain incorrect formulas, so their values are also incorrect. In addition, D3 propagates an incorrect value to D6, so its value is also incorrect. The formulas of cells B6 and D6 are correct, but could be improved by using the AVG() function, as the solution in Table 2 does. Formulas highlighted orange could be improved, but their value is correct. The feedback for each level with respect to the whole Table 1 is listed in Table 3.

|   | A | B | C | D | A | B | C | D |
|---|---|---|---|---|---|---|---|---|
| 1 | Name | Grades | | | Name | Grades | | |
| 2 | | Ex. 1 | Ex. 2 | Final | | Ex. 1 | Ex. 2 | Final |

| 3 | Anne | 92% | 58% | **17%** | Anne | 92,00 | 58,00 | =(B3-C3)/2 |
| 4 | John | 56% | 70% | **63%** | John | 56,00 | 70,00 | =(B4+C4)/2 |
| 5 | Tim | 95% | 75% | **85%** | Tim | 95,00 | 75,00 | =(B5+C5)/2 |
| 6 | Avg. | 81% | 71% | 55% | Avg. | =(B3+B4+B5) /3 | =(C3+C4+D5) /3 | =(D3+D4+D5) /3 |

*Table 1.        Example Submission Spreadsheet in Value View (Left) and Formular View (Right)*

|   | A | B | C | D | A | B | C | D |
|---|---|---|---|---|---|---|---|---|
| 1 | Name | Grades | | | Name | Grades | | |
| 2 |  | Ex. 1 | Ex. 2 | Final |  | Ex. 1 | Ex. 2 | Final |
| 3 | Anne | 92% | 58% | **75%** | Anne | 92,00 | 58,00 | =(B3+C3)/2 |
| 4 | John | 56% | 70% | **63%** | John | 56,00 | 70,00 | =(B4+C4)/2 |
| 5 | Tim | 95% | 75% | **85%** | Tim | 95,00 | 75,00 | =(B5+C5)/2 |
| 6 | Avg. | 81% | 68% | 74% | Avg. | =AVG(B3:B5) | =AVG(C3:C5) | =AVG(D3:D5) |

*Table 2.        Example Solution Spreadsheet in Value View (Left) and Formular View (Right)*

| Level | Feedback | Method |
|---|---|---|
| 1 | The spreadsheet is incorrect. | Value Matching |
| 2 | The values of cells D3, C6, D6 are incorrect. | |
| 3 | The formulas of cells D3, C6 are incorrect. | |
| 4 | The formulas of cells D3, C6 are incorrect. You should recall the info in the 'Calculating the average' tutorial. | |
| 5 | An operator of cell D3 is incorrect. | Formular Analysis |
|   | A reference of cell C6 is incorrect. | |
| 6 | The operator '+' should be used in cell D3. | |
|   | The references C3:C5 should be used in cell C6. | |
| 7 | It is preferable to use an AVG-formula in cells B6, C6, D6. | Formular Quality |

*Table 3.        Feedback from the Different Levels and their Method of Analysis for the Example Submission Shown in Table 1*

## 4      Implementation

The analyses of the different feedback levels are performed using three different methods: Value Matching (FL 1-4), Formular Analysis (FL 5-6) and Formula Quality Check (FL 7). For the implementation of these methods in our ITS, we use data dependency graphs (Hofer et al., 2013; Hofer and Wotawa, 2014) as an auxiliary tool to represent the calculation paths of a spreadsheet. In a data dependency graph each node represents a spreadsheet cell. The value within a node before the colon denotes the cell and behind the evaluated value of the cell (see Figure 1). If cell B is referenced by the formula of cell A, then there exists a directed edge from node A to node B. In contrast to Hofer et al. (2013), we refer to sources as output nodes and conversely, sinks as input nodes to run a usual depth first search (DFS) algorithm for our analysis below.

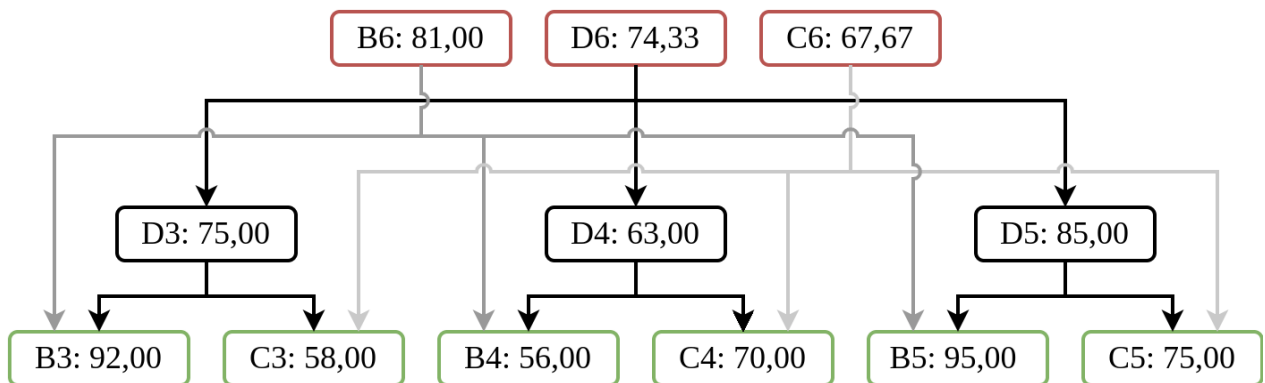

Figure 1.     Data Dependency Graph of Spreadsheet in Table 2 with Output Nodes (Red) and Input Nodes (Green)

## 4.1   Syntax Errors

Before the feedback is generated for each level, a syntax check is performed on the submitted spreadsheet. This checks if there are any syntactical errors in the given formulas that would prevent us from calculating them. If this is the case, an error will be returned instead of feedback. This is because a Spreadsheet with an error cannot be correct, and the feedback generated could be wrong.

## 4.2   Value Matching

With the data dependency graph, the values of the submission cells can be easily compared with the values of the reference solution cells in the order of their dependency on each other, even if a student has entered the information into a wrong cell. Since not all references may be present or correct in the submission, the data dependency graph of the reference solution spreadsheet is used. The following value matching analysis is used below in the FL 1-4 (with slight modifications).

1. For each output node, start a depth-first search in the graph, but only visit those parts of the graph that have not already been visited by other searches.
2. When visiting a new node, compare the solution value of the current node with the value of the corresponding submission cell. If the values do not match, mark the submission cell as a <u>value error</u>.
3. Continue with visiting the outgoing neighbors of the current node by recursively going to Step 2 for each such neighbor.
4. If all outgoing neighbors have been visited, the submission value of the current node is re-evaluated (some domain values of the children might have been corrected) and again compared to the solution value.
5. If the values still do not match, mark the submission cell as a <u>formular error</u> and replace the submission formular with the reference solution value in the spreadsheet to possibly correct a later error propagated by that cell. Otherwise, the cells formular is considered correct.
6. Finish the recursive call and return to Step 3 from which the currently visited node was called.

Using the algorithm above we present the details of the analysis for the different feedback levels.

**FL 1.** The analysis requires comparing the values of all cells in the submission to the solution. However, the analysis may terminate at Step 2 above as soon as an erroneous cell is found.

**FL 2.** All value errors and their location should be reported; thus the analysis cannot terminate early. However, Step 4 and 5 can still be omitted as no detailed information is provided at this level.

**FL 3.** Now Steps 4 and 5 are essential as these steps can distinguish between formular errors (such as D3 and C6) and (propagated) value errors (such as D6). The analysis is based on the fact that an erroneous cell, referenced in other cells, is reached and corrected (by replacing the formular with the correct value) before the referencing cells are re-evaluated. Therefore, the value of a cell referencing an erroneous cell is either correct, because the referenced cell has already been corrected, or still incorrect, which means that it is itself erroneous.

**FL 4.** In level 4, the errors found by level 3 are used and the feedback is enriched with learning material or hints, for which relevant information must be provided to the ITS. By using the column and row descriptions within the reference spreadsheet, a text classification is used to find suitable information in the learning material (Ahmad et al., 2003; Antoniu et al., 2004). Additionally, the lecturer can annotate certain cells or areas with text or links using an interactive worksheet preview.

## 4.3 Formula Analysis

In order to generate feedback for the next levels, it is not sufficient to just check the values of the cells, but their formulas must be analyzed as well. This is necessary because it is possible to get the same result with different formulas, and therefore a pure value comparison is not sufficient to determine the wrong cells. But we still need to use the previous analyses to determine which cells are incorrect. If an analysis of FL 3 marks a cell as incorrect, the following analyses are performed to generate the feedback.

**FL 5.** For level 5a, compare the functions and/or operators used between the submission and the solution. If they are the same, the ITS performs the next analysis, which generates the feedback for level 5b by using the data dependency graphs to compare the references of the submission with the references of the solution. Any differences on the children of the currently considered node between submission and solution will be returned as feedback. Otherwise, as a final analysis for FL 5, check that the constants in the Excel spreadsheet match the constants used in the solution.

**FL 6.** The analyses from FL 5 are used and the result is enriched with more information. The specific information from the solution is injected into the feedback based on the error found by the previous analyses. For example, if the submission used a different operator, the feedback would include the operator used by the solution. Vice versa, if the submission uses an operator not used in the solution, this is also reported. For example, the reference solution uses the formula *SUM(...)* and the submission uses *ROUND(SUM(...);...)*. In this case, the feedback "The function '*ROUND*' is used too often..." is returned. At this level it is also possible to recognize and report spelling mistakes.

## 4.4 Formula Quality

**FL 7.** To allow the system to not only guide the student in the creation of correct solutions but also provide feedback that guides towards the creation of high-quality solutions, which are readable, comprehensible, and therefore maintainable. Some of the metrics described in (Hofer et al., 2021) are collected as follows.

Size metrics (number of sheets, number of errors, number of cells with values/formulas) are collected during the construction of the submission data dependency graph described above. The data dependency graph is then used to determine the number of input/output cells and the most influential cells that are frequently referenced or that reference many other cells. The formulas are then analyzed for the number of operands and operators and the depth of nesting.

The metrics are collected and compared with the metrics of the reference solution, which are determined in the same manner when the solution is uploaded by the lecturer; if these metrics differ by more than a threshold defined by the lecturer, the student is given feedback that their solution could be improved, and the improvement suggestion is provided. For example, if a student uses the formula =SUM(B3:B5)/3 or =(B3+B4+B5)/3 (as in Table 1) to calculate the average, but the reference solution uses the AVG function, the student will receive feedback suggesting the use of the AVG function.

## 4.5	Feedback Presentation in the ITS

To present the feedback described above, we need a separate display for spreadsheet submissions, as shown in Figure 2. This was done because a simple text output for the display of many erroneous cells becomes very confusing. The display includes the task name, the error cells highlighted with a red box, the feedback is given, and whether the submission failed or passed. There is also the possibility to open a spreadsheet (click on the spreadsheet icon) to see the erroneous cells in the spreadsheet to see better where the cells are located.

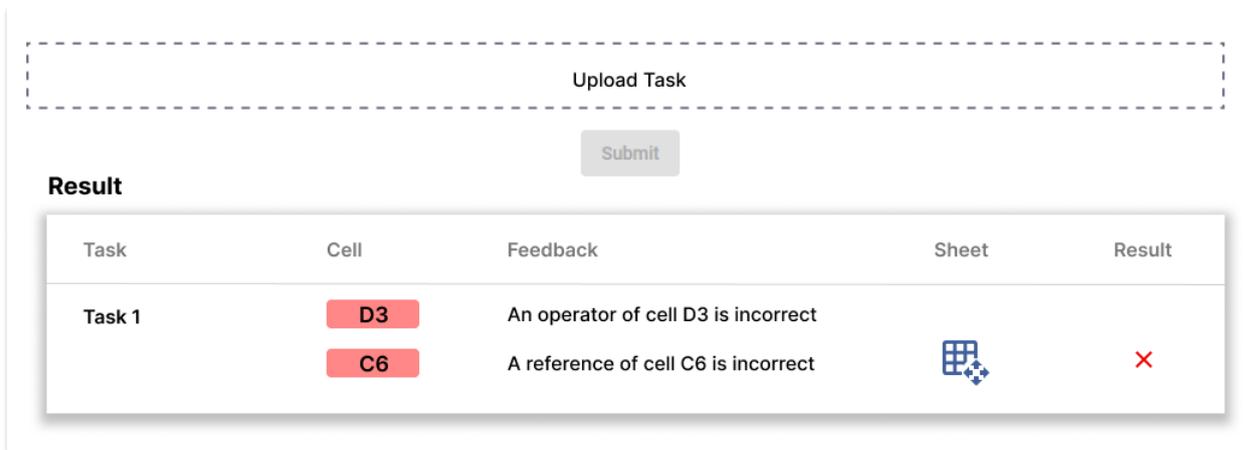

*Figure 2.	Feedback (FL 5) Displayed in the ITS*

## 4.6	Adapting Feedback Levels for SQL and Java

Although these feedback levels were initially developed for Excel tasks, they can be adapted for other types of tasks, like SQL and Java. However, comparing SQL and Java solutions to a reference implementation is much more complex since there are numerous semantically equivalent solutions, which means that solutions that produce the same result can be very different from the reference solution. Particularly with Java, the program's output or behavior cannot be compared cell by cell and formula by formula like in Excel. Therefore, we will try to discuss possible adaptation in this section.

The value matching process (FL 1-4) can be adapted for both as an output or behavioral analysis, as the results of the SQL query or Java program get compared with a reference. With SQL assignments, where DQL and DML queries are submitted for a given dataset, the output tables can be compared to those produced by a reference query, while the lecturer can specify whether the order of rows or columns should be ignored, or how missing values should be handled. To locate the source of errors within a query, several indicators can be used in the resulting table. An unequal number of rows or incorrect rows compared to the reference table indicates an error in the WHERE clause, an unequal number of columns or incorrect selection of columns indicates problems in the SELECT (or INSERT/UPDATE) clause. Errors in JOIN clauses can be detected by analyses that focus on both rows and columns. Incorrect sorting or grouping is, of course, caused by incorrect ORDER-BY and GROUP-BY clauses. For DDL statements such as CREATE, the resulting data schemas must be compared to a reference schema so that the output statement behaves the same. This approach differs from SQL Tutor in that we only compare the resulting tables and not the submitted query and therefore ensure that all queries are recognized as correct if they produce the same tables. Analogous with Java tasks, the textual output can be compared to a reference program, or if there is no output result, the behavior of the program can be analyzed through unit testing. To do this, individual methods or classes must be tested in isolation against a reference implementation. Because they are tested separately, a faulty method can be found quickly.

A semantic analysis can be used for formula analysis (FL 5-6). This needs to analyze the semantic differences with the reference and identify the necessary changes that need to be made to achieve the equivalent semantic meaning as the reference. Tools like JavAssess can be used for this in Java. It is also useful to extend the reference with a pool of references, which is always extended by correct submissions (if necessary harmonized). This will increase the number of correct submissions, and students will not be forced to design a solution that exactly matches the reference implementation.

For the quality analysis (FL 7), there are plenty of existing technologies that assess the style of database queries and computer programs (so called linting e.g., with sonarqube) that can be used for this step. In addition, benchmarking can also be an important quality measure when done correctly.

# 5    Evaluation

We conducted a study to evaluate how feedback levels affect a student's ability to correct an error using that feedback. In this study, students were asked to complete a series of tasks. Each task consists of a problem description, a spreadsheet partly solving the problem but still containing one or two errors, and feedback from one of the levels (1-3, 5-6) that assesses the spreadsheet. Every participant was asked to complete the same series of tasks: one training task designed to familiarize them with the spreadsheet interface, and four subsequent tasks that are directly relevant to the study. Each of these tasks contained an error-correcting challenge. The key variation is that for each task, individual participants were given a different, randomly assigned feedback level to work with. This means that while the task sequence remained consistent, the level of feedback received varied for every participant. We excluded FL 4 and FL 7 from our study. FL 4 was not considered because it aims for students to repeat existing learning materials and then try the task again, which does not fit the setting of the study. FL 7 was not included because the attempted solution was not developed by the students themselves and a quality assessment would be too artificial. To provide a realistic picture, the problem descriptions were taken from a previous lecture so that the tasks are solvable for the participants. For each of the two selected problem descriptions, two different incorrect student submissions were selected from the previous lecture. These partly solved solutions were then added to the corresponding problem descriptions. The participants should then use the respective feedback to correct the spreadsheet so that it solves the task. After each task, the received feedback was evaluated by the student in a questionnaire.

In this study, 62 students completed at least one task, resulting in a total of 135 completed tasks. These participants are enrolled in a variety of study programs: (applied) computer science, social media systems, economics and some were studying something else. Participants' ages ranged from 18 to 39, with an average age of 24. Regarding gender, 26 participants identified as male, 17 as female, 3 as diverse and 16 participants provided no information. Participants were also asked to rate their own experience with Excel. Of these, 24 participants rated themselves as inexperienced, 22 as moderately experienced, 12 as experienced and 4 as very experienced.

For the survey, participants were asked to evaluate the usefulness, understandability, and information quality of the feedback. Feedback is an important part of the learning process, but only as long as the student receives useful information for task completion. In particular, when feedback addresses the task or explains how to proceed with the task, it significantly impacts student learning outcomes (Hattie and Timperley, 2007). Otherwise, this can lead to student dissatisfaction and poor learning outcomes (Sadler, 1989; Nicol and Macfarlane-Dick, 2006; Keuning et al., 2016; Wong et al., 2012). Hence, we considered the following questions for the usefulness of our feedback: 1. the feedback helped me to better understand the error in the task; 2. the feedback helped me to find the error in the task faster; 3. the feedback helped me to effectively fix the error in the task.

Former studies revealed clearly feedback can be powerful (Hattie and Timperley, 2007). Therefore, the feedback must also be understandable because students are more likely to increase effort when the intended goal of the feedback is clear (Kluger and DeNisi, 1996). Unclear feedback can result in negative outcomes, engender uncertain self-images, and lead to poor performance (Thompson, 1999; Thompson and Richardson, 2001). Even more, the student might stop using the feedback if it is incomprehensible (Sadler, 1989; Nicol and Macfarlane-Dick, 2006; Keuning et al., 2016;). Hence, we asked the following questions for the comprehensiveness of our feedback: 1. the feedback was easy to understand; 2. the feedback was unambiguous; 3. the feedback was specific; 4. the feedback was difficult to understand.

Qualitative feedback information contains relevant, sufficient, accurate, and appropriate information for the students. If the provided feedback information is irrelevant or inaccurate, students may doubt the ITS's ability to generate feedback. This may affect their trust in the ITS and also negatively their learning outcome because they spend much effort on information scrutinizing (Zhou, 2013). Hence, information quality helps students to solve problems and to achieve learning objectives (Nicol and Macfarlane-Dick, 2006; Segers et al., 2008). Therefore, we considered the following questions for the information quality of our feedback: 1. the feedback contained relevant information for the task processing; 2. the feedback contained sufficient information for task processing; 3. the feedback contained detailed information for task processing; 4. the feedback contained appropriate information for task processing.

All items for the evaluation of the feedback were adapted from extant literature to improve content validity (Hair et al., 2011; Straub et al., 2004) and were measured using a five-point Likert scale ("strongly disagree" to "strongly agree"). The evaluation of the questionnaire shows that the usefulness increases with the feedback level. FL 6 is the most useful, with an average value of 3.56, and FL 1 has the lowest average value of 2.47. The evaluation of the feedback's comprehensibility shows that the feedback's comprehensibility increases with the feedback level. FL 6 is the most understandable, with an average value of 3.51, and FL 1 has the lowest average value of 2.92. The quality of information also increases with the feedback level. FL 1 achieves the lowest average value with 2.52, and FL 6 the highest with 3.63. The exact values for the various questions can be found in Figures 3 to 5.

Regarding the correctness, we checked all submissions from students that (strongly) agreed with the usefulness, understandability and of information quality of the feedback. The results revealed that the correctness increases with the FL. None of the students solved the tasks with FL 1 correctly, whereas FL 2 solved 60%, FL 3 solved 70%, FL 5 solved 75% and FL 6 solved 83% of the tasks correctly.

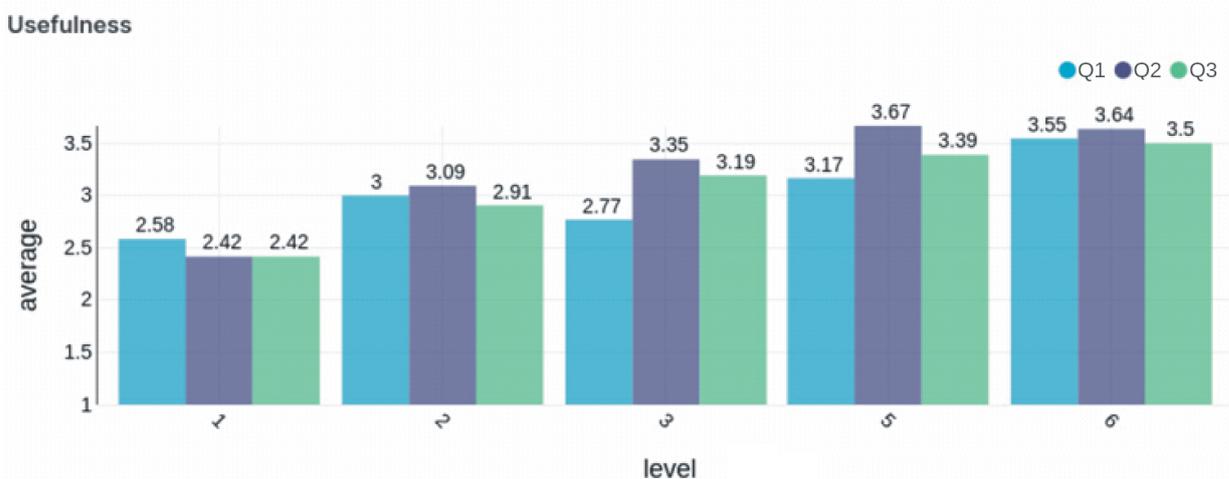

*Figure 3.     Evaluation of Usefulness. Questionnaires: Q1 the feedback helped me to better understand the error in the task; Q2 the feedback helped me to find the error in the task faster; Q3 the feedback helped me to effectively fix the error in the task.*

*Improving Automated Spreadsheet Feedback*

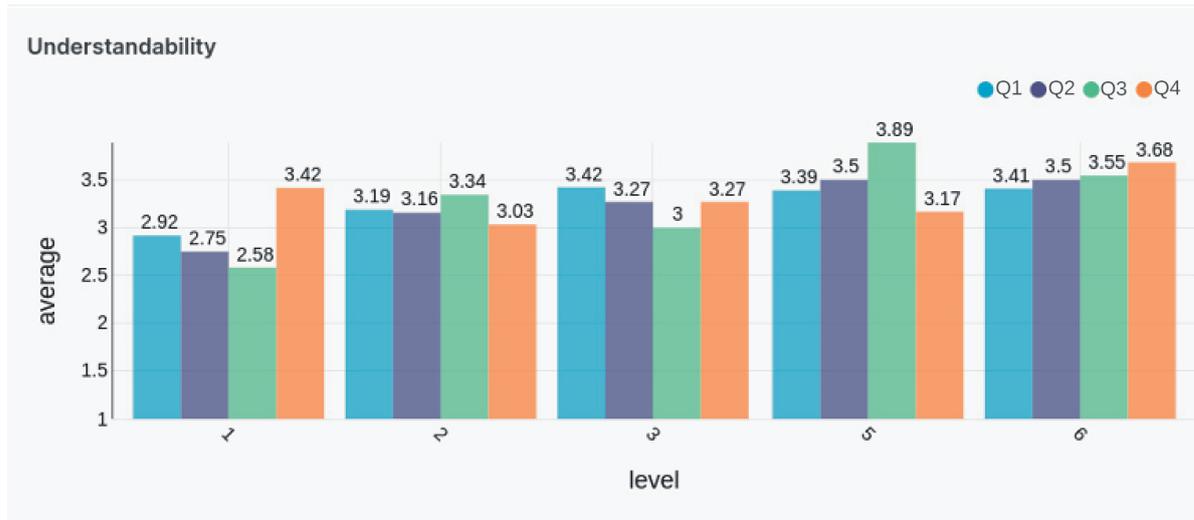

*Figure 4.   Evaluation of Understandability. Questionnaires: Q1 the feedback was easy to understand; Q2 the feedback was unambiguous; Q3 the feedback was specific; Q4 the feedback was difficult to understand.*

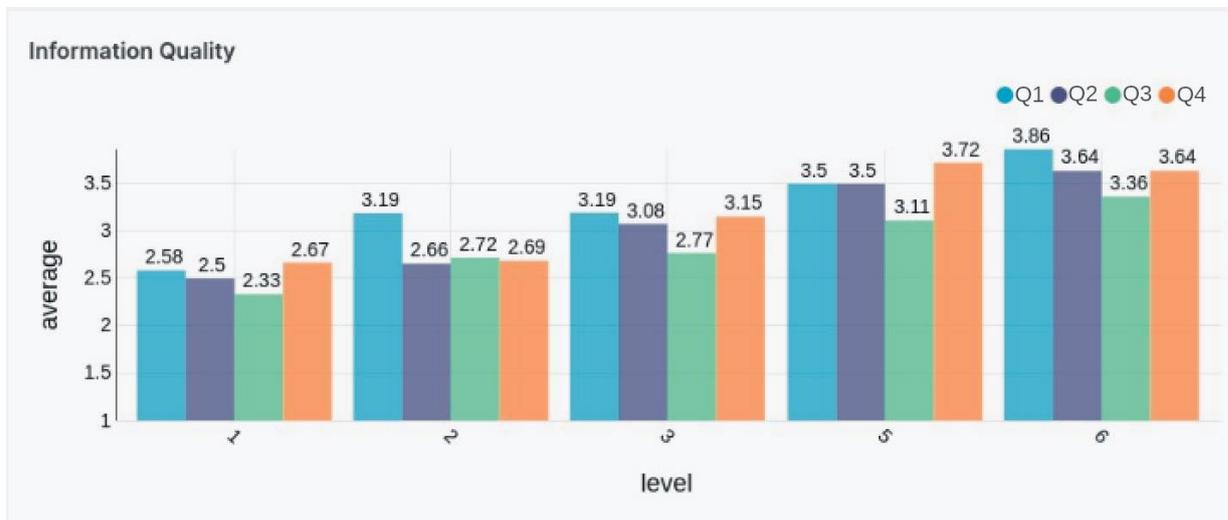

*Figure 5.   Evaluation of Information Quality. Questionnaires: Q1 the feedback contained relevant information for the task processing; Q2 the feedback contained sufficient information for task processing; Q3 the feedback contained detailed information for task processing; Q4 the feedback contained appropriate information for task processing.*

## 6      Discussion

To the best of our knowledge, this is the first study that developed an ITS for spreadsheets, which can provide fully automatic and individualized feedback based on a student submission. The feedback is generated without the lecturer having to provide feedback. The lecturer only needs to upload a reference solution to the ITS.

To optimize the learning progress of the students, we have developed seven feedback levels that build on each other. The first three levels give basic information about erroneous cells. FL 4 extends the previous level with lecture materials to review. Levels five and six give more specific hints about the concrete errors in the cells. The FL 7 is for improving the written formula's quality. These levels are designed to help the student to find the solution step by step and become a self-regulated learner (Nicol and Macfarlane-Dick, 2006). To integrate the feedback levels into our ITS, we developed analyses to generate them automatically. For this purpose, only a reference solution is needed from a lecturer. When the student uploads a submission, a data dependency graph is used to compare the submission to the reference solution and to generate feedback. For FL 1-4, a value comparison is performed based on this graph. For FL 5 and 6, the analysis compares the solution formula to the submission. The analyses for FL 7 then generate metrics based on the data collected while generating the data dependency graph.

In general, the results of the survey confirmed the importance of feedback for the learning process from the students' perspective. With respect to the usefulness of the provided feedback, the results revealed that the usefulness of the provided feedback by the ITS increases with the feedback level. The result confirms that the feedback information must be helpful to understand the error in the task. Useful feedback also increases the students' learning success and avoids frustration in learning (e.g., Nicol and Macfarlane-Dick, 2006; Keuning et al., 2016; Wong et al., 2012). Besides the usefulness of our feedback, the feedback should also be comprehensible for the students because students actively construct his or her own understanding of feedback messages (e.g., Black and William, 1998). In this case, it will be difficult for the student to apply the information when working on the task. However, the results indicated that students perceived the provided by our ITS as understandable. This holds for all feedback levels. Therefore, the provided feedback is clear and can help the students to achieve positive learning outcomes (e.g., Hattie and Timperley, 2007; Thompson 1998; Thompson and Richardson, 2001). Concerning the information quality, the survey results confirmed that the provided feedback by the ITS for all feedback levels contains relevant, sufficient, accurate, and appropriate (related to the task) information and therefore support the student's task proceeding. These results are consistent with existing studies (Nicol and Macfarlane-Dick, 2006; Segers et al., 2008; Zhou, 2013).

The correctness according to feedback levels confirms that hints are the most effective form of feedback when they help students reject erroneous hypotheses and provide direction for the solution (e.g., Earley, 1988; Harackiewicz, 1979; Harackiewicz et al., 1984; Wood and Bandura, 1987). Moreover, the results also confirm students' positive perception concerning usefulness, comprehensibility, and information quality. This means that the feedback from our ITS helps students to complete tasks and thus supports the learning process. Interestingly, former studies stressed that knowledge of results can be useful for students (e.g., Sturges, 1978; Lysakowski and Walberg, 1982; Tenenbaum and Goldring, 1989; Narciss, 2008). However, we conclude from our evaluation that for more complex tasks (e.g., spread sheets with many fields) that knowledge of results or corrective feedback does not direct students to the right way of solution.

Although the lecturer only needs to provide one reference solution, we were able to demonstrate that it is possible to evaluate spreadsheet submissions and generate fully automatically feedback in a teaching environment. Even higher feedback levels, for which more precise analyses of formulas are necessary, could be implemented in our ITS. This means that even classes with large numbers of students can be supported with individualized feedback and the feedback can be given. As for the students, the feedback levels help to support the learning process and avoid student frustration by influencing the learning process through targeted, individualized feedback. Furthermore, students can be encouraged to work out solutions independently through needs-based feedback. In addition, feedback for the students can be provided regardless of location or time, allowing them to follow their own pace of learning and working without constraints.

## 6.1    Limitations and Future Research

This study is not without limitations. First of all, concerning the feedback, FL 4 heavily depends on the learning materials provided. Although FL 4 provides the same information as FL 3 and in addition references to learning material, it is conceivable that students get confused by the learning material.

Secondly, being the basis for all of the feedback, the reference solution exerts a high impact on the FL and the learning outcome. As FL 5 and 6 give detailed hints what part of a formula is not correct, students are pushed to the reference solution which is the basis for these hints. However, in spreadsheets the same calculation can be expressed by many different formulas, so that it may happen that the feedback leads students away from their own solution. Therefore, the set of reference solutions need to be expanded because a greater variety of reference solutions is advantageous in improving students' understanding of spreadsheets error and prioritizing individualized learning.

Thirdly, the study demonstrated the usefulness of the feedback for the evaluated task by the students. However, it still needs to be investigated when which feedback level is necessary and useful depending on the difficulty of the task.

Fourthly, even so our study showed that higher feedback levels are perceived as more useful by the students, less information is better as long as the students manage to solve the tasks. The more detailed feedback is, the less students have to cogitate the solution approach. As a result, the learning outcomes may be reduced when more detailed feedback is given. It should be only avoided that students get stuck and frustrated. Thus, future research should investigate how and when FL should be revealed to the students, i.e., whether feedback of the next higher level is displayed after every $n^{th}$ incorrect submission to an assignment, whether the feedback level increases per assignment or for all assignments together, and when/if the level decreases again.

Fifthly, the number of students participating in the evaluation of the study is quite small. In addition, students faced an artificial setting where they had to correct a task which they had not started. Future research should investigate the real usage of the ITS and how it affects the learning outcome. For this, more data should be collected with additional comparison groups.

Finally, with the implemented analyses automated grading for spreadsheet exams could also be introduced and remove further work from the lecturer.